# Dirac-Source Diode with Sub-unity Ideality Factor

Gyuho Myeong[1†], Wongil Shin[1†], Seungho Kim[1], Hongsik Lim[1], Boram Kim[1], Taehyeok Jin[1], Kyunghwan Sung[1], Jihoon Park[1], Michael S. Fuhrer[2], Kenji Watanabe[3], Takashi Taniguchi[3], Fei Liu[4**], Sungjae Cho[1*]

**An increase in power consumption necessitates a low-power circuit technology to extend Moore's law. Low-power transistors, such as tunnel field-effect transistors (TFETs)[1-5], negative-capacitance field-effect transistors (NC-FETs)[6], and Dirac-source field-effect transistors (DS-FETs)[7-10], have been realised to break the thermionic limit of the subthreshold swing (SS). However, a low-power diode rectifier, which breaks the thermionic limit of an ideality factor (η) of 1 at room temperature, has not been proposed yet. In this study, we have realised a DS Schottky diode, which exhibits a steep-slope characteristic curve, by utilising the linear density of states (DOSs) of graphene[7]. For the developed DS Schottky diode, η < 1 for more than two decades of drain current with a minimum value of 0.8, and the rectifying ratio is large (> $10^5$). The realisation of a DS Schottky diode paves the way for the development of low-power electronic circuits.**

Power consumption of integrated digital devices sets the ultimate limit to downscaling and Moore's Law[11]. Reducing power consumption has been thwarted by fundamental limits on the operating voltage set by thermionic emission[12]. For an ideal thermionic device the dependence of current $I$ on voltage $V$ is expressed through the subthreshold swing SS = $[d\log(I)/dV]^{-1}$ = $(k_BT/q)\log(10)$ ≈ 60 mV/dec at room temperature, where $k_BT$ is the thermal energy and q is the elemental charge.

Two-dimensional (2D) van der Waals (vdW) materials[13,14] have been proposed for various schemes to overcome the thermionic limit (SS = 60 mV/dec) of metal-oxide semiconductor field-effect transistors (MOSFETs) in nonconventional transistors such as TFETs, NC-FETs, and DS-FETs[1-10]. In particular, DS-FETs use the linear energy dispersion relationship of graphene, producing a super-exponential change in the DOS with energy[15]. As a result, DS-FETs have achieved a smaller SS than that of a MOSFET, with a large drive current[7-10].

[1] Department of Physics, Korea Advanced Institute of Science and Technology (KAIST), Daejeon, Korea

[2] ARC Centre of Excellence in Future Low-Energy Electronics Technologies, and School of Physics and Astronomy, Monash University, Clayton, Victoria 3800, Australia

[3] National Institute for Materials Science, Namiki Tsukuba Ibaraki 305-0044, Japan

[4] Institute of Microelectronics, Peking University, Beijing, 100871, China

[†]These authors contributed equally to this work
* Corresponding author, S. C, Email: sungjae.cho@kaist.ac.kr

** Corresponding author, F. L, Email: feiliu@pku.edu.cn

Integration of heterogeneous electronic components on a single low power-consumption platform is highly desirable to enable application such as the Internet of Things. Schottky diodes are important electronic components with low operation voltage and high current[16], and have many useful applications such as rectifiers, mixers, selectors, switches, photo detectors and solar cells[16]. Although there has been considerable development of low-power transistors, steep slope diode (or diode) rectifiers that overcome the thermionic limit ($\eta < 1$) of conventional diodes have not been proposed yet, but will be necessary for device integration with low-power transistors. Herein, we propose a DS Schottky diode as an essential element for low-power circuits. The DS injects cold electrons without a long thermal tail above the potential barrier in the channel (Fig S1). Our proposed DS Schottky diode consists of a graphene/MoS$_2$/graphite heterojunction, where graphene acts as a cold electron injector, whereas the graphite/MoS$_2$ interface provides a Schottky barrier for rectification. The MoS$_2$ channel was chosen because of its high-gate tunability and mobility[17]. The minimum and average values of $\eta$ for the DS Schottky diode are 0.76 and less than 1 over more than two decades of current at room temperature, respectively, with a high rectifying ratio ($> 10^5$).

The proposed DS Schottky diode device (Fig. 1a) consists of four components: (i) an n-type monolayer MoS$_2$ channel, (ii) a graphene DS neutral at a zero gate-voltage, (iii) a graphite drain-contact to form a Schottky barrier between the graphite and monolayer MoS$_2$ for electrical rectification with a bias voltage, and (iv) metal (back and top) gate electrodes to tune the Fermi levels of 2D materials. Two-dimensional van der Waals epitaxy was performed inside an Ar-filled glovebox until the heterostructure was encapsulated by hexagonal boron nitride (hBN) to avoid any contamination through air exposure or chemicals (Fig. S2). Unlike a metal contact, a graphite contact with the monolayer MoS$_2$ forms a non-reactive clean interface[18], preventing Fermi-level pinning[19] (Fig. S3 and Fig. S4). The diode has a local top-gate and a global back-gate. The top gate only modulates the channel of the monolayer MoS$_2$ band, whereas the global back gate affects the graphene/MoS$_2$/graphite heterostructure. The gate-dependent electrical measurements (Fig. S5) indicate that the Dirac point of hBN-encapsulated graphene not on MoS$_2$ is located at $V_{BG} = 0$, whereas the Dirac point of graphene on MoS$_2$ is located at $V_{BG} \approx -18$ V because of the n-doping caused by the monolayer MoS$_2$[20].

Fig. 1b presents the characteristic drain current ($I_D$) versus bias voltage ($V_{bias}$) curve for the DS Schottky diode at $V_{BG} = -30$ V. At $V_{BG} = -30$ V, the G1 and G2 regions of graphene are p-type. When a bias voltage is applied to the graphene, electrons are injected from the p-type graphene source to the graphite drain. The electrical measurements reveal a nearly Ohmic graphene/MoS$_2$ contact and a Schottky barrier of the graphite/MoS$_2$ contact (Fig. S6). When a negative back-gate voltage is applied, the Schottky barrier height increases, and the device current is mainly modulated by the Schottky barrier at the interface between the graphite and monolayer MoS$_2$.

The performance of a Schottky diode is mainly characterised by two figures of merit. One is the rectifying ratio, which refers to the ratio between the on and off currents ($R = \frac{I_{on}}{I_{off}}$), whereas the other is $\eta$, which is the slope representing the change in drain current with a bias voltage and can be obtained from the following Schottky diode equation:

$$I_D = I_S(1 - e^{qV_{bias}/\eta k_B T}), \qquad (1)$$

where q is the elementary charge, $V_{bias}$ is the applied bias voltage, η is the ideality factor, $k_B$ is the Boltzmann constant, T is the temperature, and $I_D$ and $I_S$ are the drain and leakage currents, respectively. Eqn (1) corresponds to SS = $(\eta k_B T/e)\log(10)$ hence values η < 1 correspond to SS below the thermionic limit. The characteristic curve at a negative gate voltage in Fig. 1b exhibits rectification behaviour with η < 1 observed over more than two decades of drain current, a minimum η of 0.76, and a large rectifying ratio (> $10^5$).

To explore the switching mechanism of DS Schottky diode, we developed analytical formula for ideality factor and performed numerical device simulations (See supplementary materials 7). Both the two methods show that the ideality factor less than 1 is obtained in DS diode due to the linear density of states of graphene. The switching slope of a diode is determined by the energy-dependent current density injected from an electrode, which is related to DOS and the distribution function. Graphene has a linear energy-dependent electronic DOS near the Dirac point, which is different from conventional metals with a constant DOS around the Fermi level. Therefore, the thermal tail of the Boltzmann distribution function is suppressed by the Dirac point tuned to the off-state region by doping. Namely, as the bias voltage is decreased on the graphene electrode as shown in Fig. 1c, the part of current density related to the distribution function is increased exponentially similar as conventional metals, which results in the ideal factor limit of 1. While, the injected DOS over the top of channel barrier is also increased linearly from off-state to on-state, as shown in Fig. 1c. Therefore, current is increased super-exponentially and the ideal factor below 1 is obtained in the diode with graphene electrode as the injection source.

Therefore, the switching slope of a diode, i.e., η < 1, is obtained in the diode with a graphene electrode as the cold electron injection source because of the linear DOS of the DS. Detailed simulation results are presented in Fig. S7. Quantum transport simulations show that DS diode has promising device performance. The ideality factor as small as 0.69 is obtained in the simulated DS diode and is less in 1 in more than five decades of current at room temperature.
The on-state current is larger than $10^3$ μA/μm and the rectifying ratio is over $10^7$.

Fig. 2a presents the $I_D$-$V_{bias}$ characteristic curve of the DS Schottky diode at different back-gate voltages. For the proposed DS Schottky diode to work as a diode, an asymmetric Schottky barrier height between the source and drain is necessary[21-24]. To satisfy this condition, we placed asymmetric graphene and graphite contacts with the monolayer $MoS_2$ channel with gates. Without gate modulation, graphene has a work function of 4.3–4.7 eV from a monolayer to a few layers[25-27]. Because the work function of graphene (~4.3 eV) does not differ significantly from the electron affinity of $MoS_2$ (~4.2 eV)[28-31], the Schottky barrier height at the graphene/$MoS_2$ interface is negligible, compared to the Schottky barrier height at the graphite/$MoS_2$ interface. This also indicates that the Dirac point of pristine graphene is located near the conduction band edge of $MoS_2$. Fig. S6 indicates that the graphene/$MoS_2$ device shows an almost Ohmic IV curve, whereas graphite/$MoS_2$ does not show an Ohmic IV curve at room temperature. Fig. 2a shows that as the gate voltage decreases, the rectification behaviour becomes dominant at negative gate voltages. As the back-gate voltage exceeds $V_{BG} > 0$, non-diode $I_D$-$V_{bias}$ characteristic curves appear.

To clarify the origin of the gate-dependent modulation of the $I_D$-$V_{bias}$ characteristic curves, we measured the modulation of the Schottky barrier height with back-gate voltages from the activation energy in the reverse bias regime. The Schottky diode equation (Eq. 1) can be rewritten as

$$I_D = AA^*T^\alpha e^{-q\Phi_B/k_BT}\left(1 - e^{\frac{qV_{bias}}{\eta k_B T}}\right), \quad (2)$$

where $A$ is the area of the Schottky junction, $A^*$ is the Richardson constant, $\alpha = 3/2$ is an exponent for a two-dimensional semiconducting system[32], $k_B$ is the Boltzmann constant, q is the elementary charge, T is the temperature, and $\Phi_B$ is the Schottky barrier height. When a large negative bias in absolute value is applied, i.e., $e^{qV_{bias}/k_BT} \approx 0$, the saturated drain current is proportional to $T^{3/2}e^{-q\Phi_B/k_BT}$. The inset of Fig. S4a shows a plot of $\ln(I_{sat}/T^{3/2})$ versus $1/k_BT$ in the reverse bias saturation regime ($V_{bias} = +1$ V). We extract $\Phi_B$ for a given $V_{BG}$ from the slope of each curve. Fig. S4a shows the Schottky barrier height obtained from the slope of each curve in the inset of Fig. S4a. As shown in Fig. S4b, in the highly positive $V_{BG}$ regime, the device shows an almost linear $I_D$-$V_{bias}$ curve, exhibiting nearly Ohmic contact behaviour (negligible Schottky barriers on both sides of the contacts, graphene and graphite with $MoS_2$). The adjustable Schottky barrier height with gate voltage indicates that Fermi-level pinning does not exist at the interface between graphite and monolayer $MoS_2$. This absence of Fermi-level pinning is owing to the defect- and disorder-free interface between two-dimensional materials, graphite and monolayer $MoS_2$[18,19].

To prove that the proposed diode is operated via cold carrier injection from a graphene DS at negative back-gate voltages, we measured the SS to determine if it showed sub-thermionic values. Fig. S8a shows the characteristic $I_D$ versus top-gate voltage ($V_{TG}$) transfer curve under the working conditions of the DS-FET, i.e., $V_{BG} < -18$ V, where both the G1 and G2 regions of graphene are p-type. When we apply $V_{BG} = -20$ V, graphene regions G1 and G2 become heavy and slightly p-type, respectively. When the top gate placed on the $MoS_2$ channel is swept from the off state to the on state, the DOS of the graphene increases according to the band diagram presented in Fig. S8c, thereby operating as a DS-FET. As shown in Fig. S8b, the SS value of the device is less than 60 mV/dec, which indicates that the proposed diode acts as a DS-FET owing to the linear energy dispersion relationship of the graphene-source electrode, resulting in a super-exponential change in the DOS. Both DS-FET and DS Schottky diode have the same origin for SS < 60 mV/dec and $\eta < 1$.

Fig. 3 shows the $I_D$-$V_{bias}$ characteristic curve in the steep-slope diode regime at $V_{BG} = -15, -30,$ and $-45$ V, where the graphene is p-doped. At $V_{BG} = -15$ V, region G1 is p-type and region G2 is slightly n-type doped. However, as the negative bias voltage is applied, the top of the Schottky barrier is located at a valence band of graphene region G2. At $V_{BG} < -15$ V, both regions G1 and G2 are p-type. In all the regimes at $V_{BG} = -15, -30,$ and $-45$ V, where the top of the Schottky barrier is located below the Dirac point of graphene regions G1 and G2, $\eta$ of the device is less than 1 in more than two decades of current owing to the cold charge injection from the DS at a forward bias ($V_{bias} < 0$). The minimum $\eta$ that we measured in one decade of current is 0.76.

The red dotted line in Fig. 3 is an ideal diode curve ($\eta = 1$) in the forward bias direction. The DS diodes in these gate-voltage regions show rectification ratios exceeding $10^5$ at $V_{BG} = -15$ V (more than $10^4$ when $V_{BG} = -30$ V and more than $10^3$ when $V_{BG} = -45$ V). We note that the device leakage current level is limited by the leakage currents (~ 50 pA) from the measurement equipment (Fig. S9b). Therefore, the reverse bias leakage current level from the diode should be lower than the measured values.

In conclusion, we successfully demonstrated the first DS Schottky diode that operates based on cold charge injection from a graphene source owing to the linear DOS and a Schottky barrier at the interface between graphite and monolayer $MoS_2$. As the linear DOS of the injected charges from p-type graphene over the top of the Schottky barrier between graphite and n-type monolayer $MoS_2$ increases linearly from reverse to forward bias, an ideal factor below 1 is obtained in the diode with a graphene electrode as the injection source. Using gate modulation of the Schottky barrier height of the graphite/$MoS_2$ junction, gradual switching between the diode and non-diode behaviours was also observed. The fabricated DS Schottky diode presents a minimum $\eta$ as low as 0.76 in one decade of current, and it remains less than 1 for more than two decades of current at room temperature, with a high rectifying ratio exceeding $10^5$. Additionally, the device shows SS < 60 mV/dec for the same origin as that for $\eta < 1$. The realisation of a steep-slope DS Schottky diode paves the way for the development of low-power circuit elements and energy-efficient circuit technology.

## Methods

**Device fabrication**

We first prepared monolayer $MoS_2$, graphene, graphite, and hBN flakes on a 90 nm Si/$SiO_2$ wafer via mechanical exfoliation from bulk crystals in an Ar-filled glove box (< 0.1 ppm of $H_2O$ and $O_2$) to maintain clean surfaces and prevent contamination from air exposure. Monolayer $MoS_2$ and graphene were identified using the optical contrast and Raman spectroscopy. Each piece was picked up using a standard dry transfer method with a polydimethylsiloxane (PDMS) stamp covered with a polycarbonate (PC) film and transferred onto a 285-nm wafer. The PC film was washed with chloroform, acetone, and isopropyl alcohol (IPA). Then, standard e-beam lithography and $CF_4$/$O_2$ plasma etching, followed by e-beam deposition, were used to place electrical contacts on the vdW layers. Additional e-beam lithography and deposition were performed to place the gate electrode (Fig. S2).

**Measurement**

To obtain a transfer curve, we performed DC measurements from room temperature to high temperatures in a home-built measurement vacuum chamber. Yokogawa 7651 and Keithley 2400 were used to bias the DC voltages to the source and gate electrodes. A DDPCA-300 preamplifier was used to amplify the drain current ($\times 10^6$) and convert it to a voltage, and this signal was measured using a Keithley 2182a nanovoltmeter (Fig. S9a).

## Acknowledgments

We thank J. Lee for helpful discussions. S. C. acknowledges support from Korea NRF (Grant Nos. 2019M3F3A1A03079760, 2020M3F3A2A01081899, and 2020R1A2C2100258).F.L acknowledges support from NSFC (Grant No. 61974003) and the 111 Project (Grant No. B18001). M.S.F. acknowledge support from the ARC (CE17010039).

## Author contributions

S. C. conceived and supervised the project. G.M and W.S. fabricated devices and performed measurements. K.W. and T.T. grew high-quality hBN single crystals. S.K., J. P., K. S., H.L., B.K., and T.J. assisted high-temperature transport measurements. F.L. developed the theoretical model and performed device simulations. S. C, G.M., W.S., M.S.F., and F.L. analyzed the data. S.C. and G.M. wrote the manuscript. All the authors contribute to editing the manuscript.

## Competing interests

The authors declare no competing financial interests.

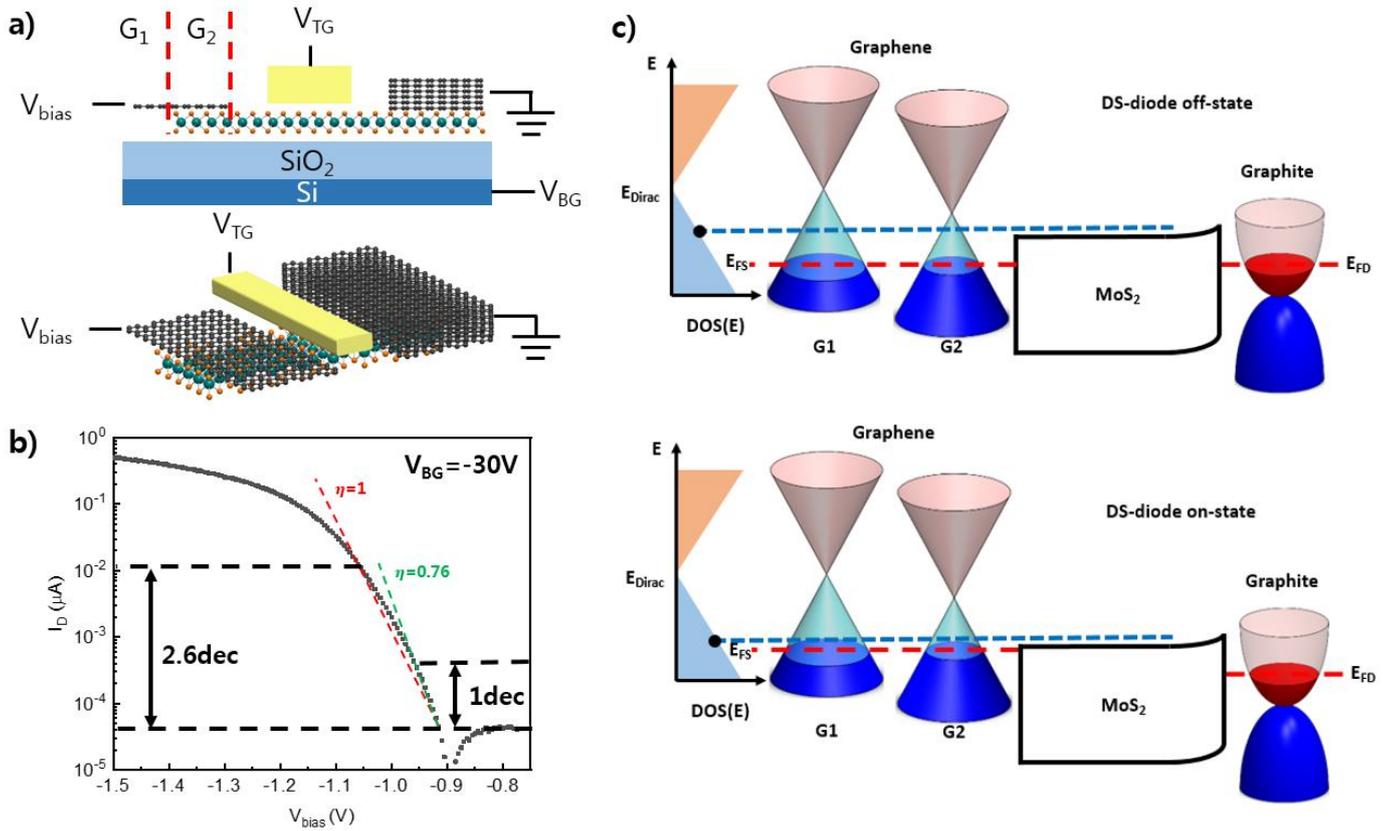

**Figure 1. a** Schematic image of graphene/MoS$_2$/graphite heterojunction diode. We used graphene as a source and graphite as a drain. The graphene source can be divided into two regions, G1, which is outside MoS$_2$, and G2, which is on MoS$_2$. **b** Characteristic $I_D$-$V_{bias}$ curve in our device, which exhibits η = 0.76 in 1 decade of current and an average η < 1 in more than two decades of current. The rectifying ratio of our device is larger than $10^5$. **c** Band diagram of DS Schottky diode, which explains the working principle of cold electron injection from graphene.

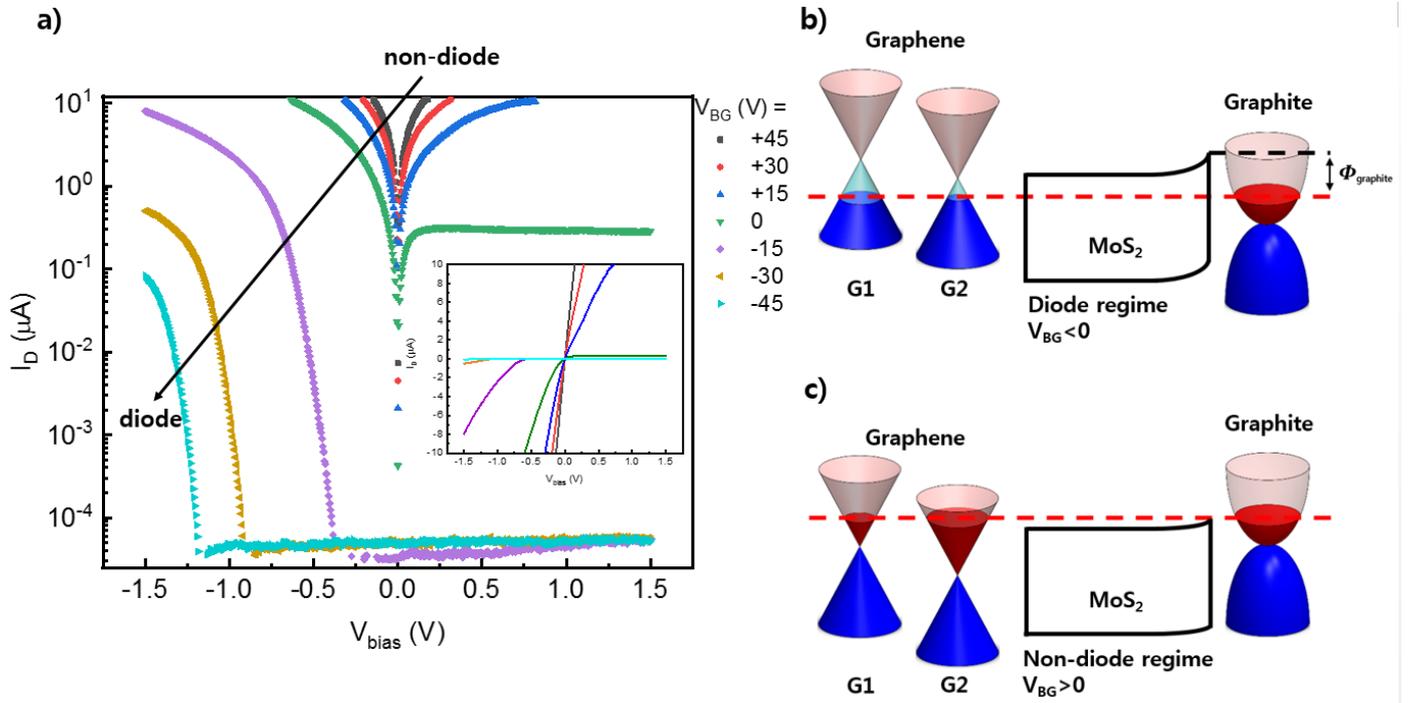

**Figure 2. Characteristic $I_D$-$V_{bias}$ curve for various $V_{BG}$ and its band diagram. a** Characteristic $I_D$-$V_{bias}$ curve in the range of $V_{BG}$ = –45 ~ +45 V. As $V_{BG}$ decreases, change from non-diode to diode behaviour is observed. **b** Band diagram when $V_{BG}$ < 0 (diode regime). Owing to the larger work function of graphite than that of graphene, the device becomes a graphite/MoS$_2$-interface Schottky barrier-dominant Schottky diode. **c** Band diagram when $V_{BG}$ > 0 (non-diode regime). Owing to the weak Fermi-level pinning between the 2D metal and MoS$_2$, the Schottky barrier height of the graphite/MoS$_2$ interface can be modulated. As $V_{BG}$ increases, the work function of graphite decreases, and the Schottky barrier height of the graphite/MoS$_2$ interface decreases.

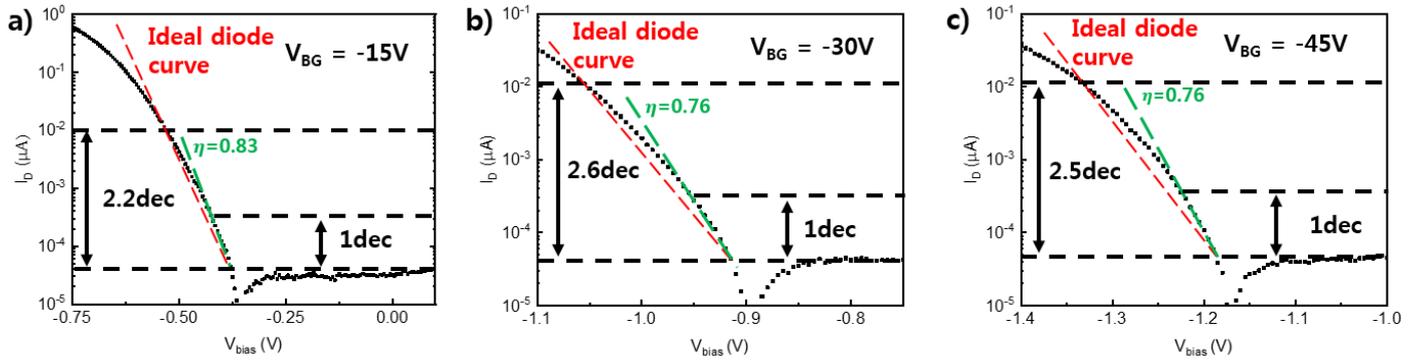

**Figure 3. Slopes of DS Schottky diode versus ideal diode and recorded ideality factor in 2D vdW material-based diode.** Comparison of slopes between the proposed DS Schottky diode and an ideal diode. Black and red dotted data represent those of the DS Schottky diode and an ideal diode, respectively. The DS Schottky diode exhibits an average η of 1 for 2.2, 2.6, and 2.5 decades when $V_{BG}$ = –15, –30, and –45 V, respectively.

**Dirac-source Schottky Triode for Low-power Electronic Circuits**

Gyuho Myeong[1†], Wongil Shin[1†], Seungho Kim[4], Hongsik Lim[1], Boram Kim[1], Taehyeok Jin[1], Kyunghwan Sung[1], Jihoon Park[1], Michael S. Fuhrer[5], Kenji Watanabe[6], Takashi Taniguchi[3], Fei Liu[4**], Sungjae Cho[1*]

[4] Department of Physics, Korea Advanced Institute of Science and Technology (KAIST), Daejeon, Korea

[5] ARC Centre of Excellence in Future Low-Energy Electronics Technologies, and School of Physics and Astronomy, Monash University, Clayton, Victoria 3800, Australia

[6] National Institute for Materials Science, Namiki Tsukuba Ibaraki 305-0044, Japan

[4] Institute of Microelectronics, Peking University, Beijing, 100871, China

[†]These authors contributed equally to this work.
* Corresponding author, S. C, Email: sungjae.cho@kaist.ac.kr

** Corresponding author, F. L, Email: feiliu@pku.edu.cn

# Supplementary Texts

## 1. Electron density distribution of normal metallic and Dirac sources

The continuous density of states (DOS) of a normal metallic source exhibits the Boltzmann distributed electron density given by $n(E) \approx \exp((E_F - E)/k_B T)$. For a Dirac source, the linearly varied DOS produces a super-exponentially decreasing electron density (given by $n(E) \approx (E_{Dirac} - E) \exp((E_F - E)/k_B T)$) with decreasing energy. Fig. S1 compares the electron densities of the normal metallic and Dirac sources. The green dotted curve in Fig. S1(b) indicates the electron density of the normal metallic source to facilitate its comparison against that of the Dirac source.

## 2. Device Fabrication

Fig. S2 illustrates the fabrication of the Dirac-source (DS) Schottky triode. As can be seen, the first step involves the preparation of a polydimethylsiloxane (PDMS) stamp covered with a polycarbonate (PC) film on a glass slide. Subsequently, $MoS_2$ flakes are mechanically exfoliated on a $Si/SiO_2$ wafer. In this study, the $MoS_2$ exfoliation was performed in an Ar-filled glove box to prevent contamination. Using the standard dry-transfer method, each flake is picked up in the order—top hexagonal boron nitride (hBN), graphite, graphene, $MoS_2$, and bottom hBN. After fabrication of the PC film and confirming sufficient adherence of the prepared flakes, the wafer was slowly heated to 90 °C, during which time, the slide glass is slowly raised. During the pick-up process, owing to the large area of the top hBN, graphene, graphite, and $MoS_2$ do not directly touch the PC film. After fabrication of the heterostructure on the PC film, the latter is slowly placed onto a prepared 285-nm-thick $Si/SiO_2$ wafer. Subsequently, the wafer is heated to 180 °C, thereby melting the PC film. Thereafter, the PC film is successively washed using chloroform, acetone, and isopropyl alcohol (IPA). After transfer of the heterojunction to a new wafer, the device is exposed to chemicals to erase the released PC film. However, graphene, graphite, and $MoS_2$ layers are encapsulated within large areas of the top and bottom hBN layers, which the chemicals cannot percolate. After fabricating the heterostructure on a 285-nm $Si/SiO_2$ wafer, the standard e-beam lithography and plasma etching procedures are performed via e-beam deposition (Cr/Au= 5/60 nm) to place electrical contacts in the graphene and graphite layers. The hBN and graphite layers are etched using $CF_4/O_2$ and $Ar/O_2$, respectively. Additional e-beam lithography and deposition processes are performed to facilitate top-gate placement.

## 3. Contact quality between normal and two-dimensional (2D) metals

With regard to the fabrication of electronic devices using transition metal dichalcogenides (TMD) materials, several recent studies have reported the improvement of the metal–semiconductor interface as a means for enhancing device performance. For example, the use of the 2D-semimetal, pre-patterned metal contacts, as well as the method of releasing evaporated metal on a TMD substrate have been suggested as ways to improve the metal–semiconductor interface[1-3]. These methods are intended to prevent the damage caused to the TMD material during the direct metal deposition. Fig. S3 illustrates the schematic process of establishing the metal-evaporation and 2D-metal contacts. The metal-deposition process comprises three steps: (1) coating polymethyl methacrylate (PMMA) resist, pattern lithography and development, (2) metal deposition, and (3) lift-off. Each step is influenced by a factor that limits the realizable contact and device quality. During lithography pattern development, the region of metal deposition is exposed to various chemicals (resistive coating, developer, and stopper) and air. During metal deposition, hot metal vapor adheres to the contact region, thereby causing structural deformations within the TMD material[1]. During lift-off, the channel region of the device is exposed to a solvent. Thus, electronic devices fabricated using direct metal deposition are subjected to the occurrence of chemical contamination and structural deformation in both the metal-contact and channel regions. This severely impacts the device performance. Contrastingly, in devices that incorporate 2D van der Waals (vdW) metal contacts, it is possible to maintain an ultraclean interface between the TMD material and metal.

## 4. Schottky barrier measurement

Owing to the weak Fermi-level pinning between the 2D metal and $MoS_2$[4], the Schottky barrier height can be altered by modifying the applied gate voltage. Fig. S4(a) depicts the measured Schottky barrier heights. Owing to limitations of the measurement system used in this study, measurements could only be performed in the voltage range of $V_{BG}$ = -12 V to 16 V. At the positive end of the applied $V_{BG}$ values, the proposed device exhibits a linear I–V characteristic, thereby indicating a near-Ohmic contact. Accordingly, an increase in $V_{BG}$ causes a reduction in the graphite and graphene work functions[5]. This decreases the Schottky barrier heights of the graphene–$MoS_2$ and graphite–$MoS_2$ interfaces.

## 5. Electron doping of graphene on $MoS_2$

A previous study has explained the doping mechanism of graphene on $MoS_2$[6]. The placement of graphene on n-type $MoS_2$, causes the former's Fermi level to become n-doped. To estimate the Fermi level of the G2 graphene region, we performed appropriate measurements over several layers. Figs. S5(a) and S5(b) depict the actual image and schematic of the test device, which was fabricated by aligning MoS₂ and graphene such that they overlapped. Therefore, the resulting transfer curves differ when measurements are performed on the "graphene side" and "across MoS₂." Figs. S5(c) and S5(d) reveal that when measurements are performed on the "graphene side," graphene exhibits a single Dirac point around $V_{BG} = 0$ V. In contrast, when measurements are performed "across MoS₂" a Dirac point is observed around $V_{BG} = 0$ V and a shoulder peak is observed around $V_{BG} = -18$ V. This shoulder peak indicates that the graphene on MoS₂ is n-doped. The electron doping of graphene on MoS₂ was confirmed by performing the "across MoS₂" measurements by sweeping the top gate placed atop the MoS₂–graphene overlap. As observed, graphene exhibits a Dirac point around $V_{TG} = -0.8$ V, which nearly equals the location of the shoulder peak depicted in Fig. S5(c).

## 6. Ohmic and non-Ohmic contact behaviors of graphene and graphite electrodes on MoS₂

Fig. S6 illustrates the contact behaviors of the graphene–MoS₂–graphene and graphite–MoS₂–graphite heterostructures. As reported previously[7], the graphene contact demonstrates a nearly linear I–V curve, which represents Ohmic behavior. In contrast, the graphite contact demonstrates a nonlinear I–V curve and non-Ohmic behavior.

## 7. Theoretical backgrounds of DS diode

### 7.1 Analytical formula for ideality factor less than one of DS diode

The Dirac point of graphene is above the conduction band edge of MoS₂ as show in Fig. 1c, and the doping of graphene is p-type at a negative gate voltage. When the bias voltage applied to graphene decreased, the injected current density increased differently from that in the case of normal metals. The current is described using Landauer approach, as follows:

$$I = \frac{2q}{h} \int_{-\infty}^{\infty} dE\, T(E) M(E) [f(E - E_{F,S}) - f(E - E_{F,D})], \quad (1)$$

where T(E) is the transmission at energy E, M(E) is the number of channels at energy E, f(E) is the Fermi distribution function of the contact, and $E_{F,S}$ and $E_{F,D}$ are the Fermi levels of the source and drain, respectively. In this case, $E_{F,S} = 0$, $E_{F,D} = E_{F,S} - qV_{bias} = -qV_{bias}$. Because graphene has a linear density of state, $M(E) = M_0|E - E_D|$, where $E_D$ is the Dirac point of the graphene source. The thermionic current has a transmission, $T(E) = 1$. Applying the above conditions to Equation 1, we obtain:

$$I = \frac{2qM_0}{h} \int_{-\infty}^{\infty} dE |E - E_{Dirac}|[f(E) - F(E + qV_{bias})]. \quad (2)$$

When the Dirac point of graphene is above the top of barrier of MoS$_2$, the energy between the Dirac point of graphene and the top of barrier of MoS$_2$ is larger than a few $k_B T$.

$$I = \frac{2qM_0}{h}\left(1 - e^{-\frac{qV_{bias}}{k_B T}}\right)\left\{\int_{E_{top}}^{E_D} dE(E_D - E)e^{-\frac{E}{k_B T}} + \int_{E_D}^{\infty} dE(E - E_D)e^{-\frac{E}{k_B T}}\right\}$$

$$= \frac{2qM_0 k_B T}{h}\left(1 - e^{-\frac{qV_{bias}}{k_B T}}\right)\left[2k_B T e^{-\frac{E_D}{k_B T}} - (E_{top} + k_B T - E_D)e^{-\frac{E_C}{k_B T}}\right]$$

$$\sim \frac{2qM_0 k_B T}{h}\left(1 - e^{-\frac{qV_{bias}}{k_B T}}\right)[E_D - E_{top} - k_B T]e^{-\frac{E_C}{k_B T}}$$

$$= J_0 \left(1 - e^{-\frac{qV_{bias}}{k_B T}}\right)[E_D - E_{top} - k_B T], \quad (3)$$

where $E_{top}$ is the top of barrier of MoS$_2$.

From Equation 4, η can be defined as

$$\eta = -\frac{q \partial V_{bias}}{k_B T \partial \log I_D} = \frac{1}{1 + \frac{k_B T}{E_D - E_{top} - k_B T}} \quad (4)$$

and when $E_D - E_{top} = nk_B T$,

$$= \frac{1}{1 + \frac{1}{n-1}} = \frac{n-1}{n} < 1, \quad (5)$$

i.e., η becomes less than 1.

### 7. 2 Theoretical calculations and simulations of DS diode

Quantum transport simulations are performed to explore switching properties of DS Schottky diode. The DS diode consists of graphene, monolayer (ML) MoS$_2$ and graphite vdW heterojunction with graphene as a cold electron source. Both graphene and ML MoS$_2$ are described by a Dirac Hamiltonian[8]:

$$\widehat{H}_0 = at(\tau k_x \hat{\sigma}_x + k_y \hat{\sigma}_y) + \frac{\Delta}{2}\hat{\sigma}_z \quad (6)$$

Where, a is the lattice constant, t is the intralayer hopping energy and $\Delta$ is the band gap which is 0 eV for graphene and 1.65 eV for ML MoS$_2$ [8]. $\hat{\sigma}$ denotes Pauli metrics on the Pseudo spin basis. The van der Waals coupling parameters between graphene and ML MoS$_2$ are obtained by fitting the band structure of graphene-MoS$_2$ heterojunction calculated by using density functional theory (DFT). Graphite is modeled by n-type bilayer graphene Hamiltonian to simplify calculations[9]. The ballistic transport properties of DS diode are calculated by solving the Schrodinger equation within the non-equilibrium Green's function (NEGF) and Poisson's equation self-consistently[8].

Fig. S7 shows the device structure and transport properties of the simulated DS diode. The device consists of 10 nm graphite, 32 nm MoS$_2$ and 8 nm graphene. An Ohmic contact is used between p-type graphene and 7 nm n-type MoS$_2$ as shown in Fig, S7(a). The Schottky barrier between graphite and MoS$_2$ is set at 0.48 eV. The contact types are consistent with the fabricated DS diode. The top gate length is 20 nm and bias voltage is applied on p-type graphene. MoS$_2$ under the gate has a flat band at V$_G$= 0 V. It is shown that a steep slope switching is realized in DS diode at V$_G$ = 0.1 V in Fig. S7(b) and the ideality factor can be as small as 0.69 at -0.15 V < V$_{bias}$ < -0.10 V, which breaks the switching limit of conventional Schottky diode. At V$_{bias}$ = -0.05 V, the Dirac point of graphene is below the top of barrier of MoS$_2$ as shown in Fig. S7(c). As bias voltage is decreased to -0.15 V, the Dirac point of graphene gets larger than the Schottky barrier at graphite-MoS$_2$ interface as shown in Fig. S7(d). The injected carrier around the Dirac point is effectively filtered as calculated current density in Fig. S7(d). Due to the linear density of states of graphene, the current is increased super-exponentially and ideality factor is less than 1 as expected. The current is increased by over 5 orders of magnitude with the average ideality factor about 0.88 as bias voltage is decreased from -0.10 V to -0.40 V. Besides promising steep slope switching, DS diode also have high on-state current over 10$^3$ μA/μm and large rectifying ratio over 10$^7$.

## 8. Dirac-Source FET measurement

Fig. S8 shows the Dirac-source FET measurement to prove that the cold electron injection from the graphene source is operating mechanism of proposed DS diode.

## 9. Device dimensions, measurement setup, and system leakage level

Fig. S9 depicts the device dimensions, measurement setup schematic, and system leakage level. The contact areas of the graphene–MoS$_2$ and graphite–MoS$_2$ interfaces equal $21.9\ \mu m^2$ and $19.7\ \mu m^2$, respectively. The length of the channel between graphene and graphite equals 3.35 μm. The graphene and graphite contacts were used as the source and drain electrodes, respectively. Using the Yokogawa 7651 DC source, a bias voltage was applied to the graphene electrode, and the current signal from the graphite electrode was amplified to 10$^6$ times in the DDPCA-300 pre-amplifier and the corresponding voltage was extracted. The current value was measured by monitoring the signal changes between DDPCA-300 pre-amplifier and its conversion to voltage using the Keithley 2182A Nanovoltmeter. The leakage level of the proposed system approximately equals 50 pA.

## 10. Thickness characterization of MoS$_2$ and graphene via Raman spectra analysis

Fig. S10 depicts the Raman spectra of MoS$_2$ and graphene used in the proposed DS triode. As can be seen, the MoS$_2$ spectra are characterized by the E$^1_{2g}$ (in-plane) and A$_{1g}$ (out-of-plane) peaks. In this study, the thickness characterization of MoS$_2$ was performed by comparing the peak-to-peak distance between the E$^1_{2g}$ and A$_{1g}$ modes[10]. In the MoS$_2$ case, this distance equaled approximately 18.8 cm$^{-1}$, thereby indicating the use of an MoS$_2$ monolayer in the device. Meanwhile, the Raman spectra of graphene are characterized by the G and 2D peaks, and the graphene-layer thickness can be determined by evaluating the intensity ratio of the G and 2D modes, i.e., I$_{2D}$/I$_G$[11]. As observed, the graphene monolayer yields an I$_{2D}$/I$_G$ value that exceeds unity. This result is consistent with I$_{2D}$/I$_G$ value determined using the Raman spectra of graphene, i.e., I$_{2D}$/I$_G \approx 3$.

## 11. Non-zero threshold voltage in diode behavior regime

Fig. S11 depicts the mechanism of non-zero threshold voltage in diode regime. A conventional Schottky diode consisting of n-type semiconductor and metal, where Fermi level of the semiconductor is fixed near the conduction band, should have sign change of the IV curve at zero bias voltage. However, the Fermi level of the monolayer MoS$_2$ in our triode can change with the gate voltage. At V$_{BG}$ = 0V, where the Fermi level of MoS$_2$ lies near the conduction band, the IV curve of our triode shows sign change at zero bias voltage. However, At V$_{BG} \leq$ - 15 V, where the Fermi level of MoS$_2$ moves away from the conduction band with decreasing gate voltage, the sign change of the IV curves no longer occurs at zero bias voltage. We believe this is an artifact of a small (positive) offset current in the measurement setup, which overwhelms the extremely low forward-bias (negative) current over a range of negative bias voltages, as explained below.

When the Fermi level of MoS$_2$ is near the conduction band edge at V$_{BG}$ = 0 V, small negative (forward) bias applied to graphene can activate carriers that can pass over the Schottky barrier formed at the interface between graphite and MoS$_2$. However, When the Fermi level is located deep inside the bandgap, large negative bias voltage should be applied for the thermally activated electrons to pass over the Schottky barrier as in Fig. S11. The triode thus remains in the off-state (where we observe only the leakage current level of the equipment $\approx$ 50 pA) not only in the reverse bias regime, but also in the low forward bias regime where large enough bias is not applied so that the carriers cannot pass over the Schottky barrier.

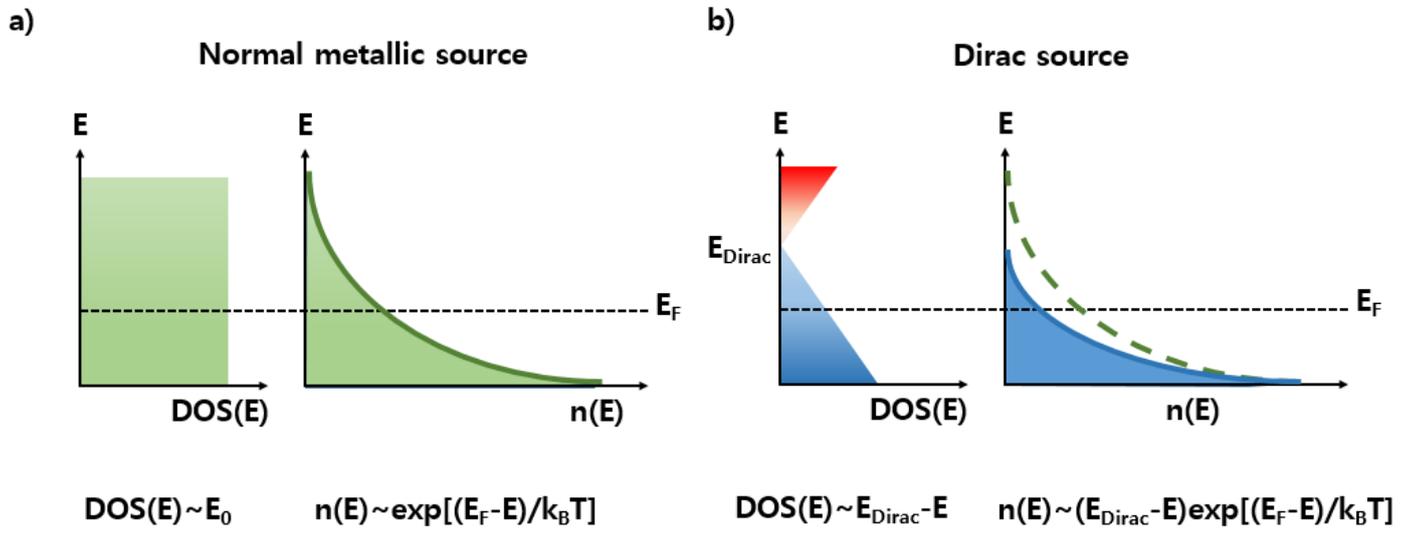

**Figure. S1** State and electron densities of (a) normal metallic and (b) Dirac sources.

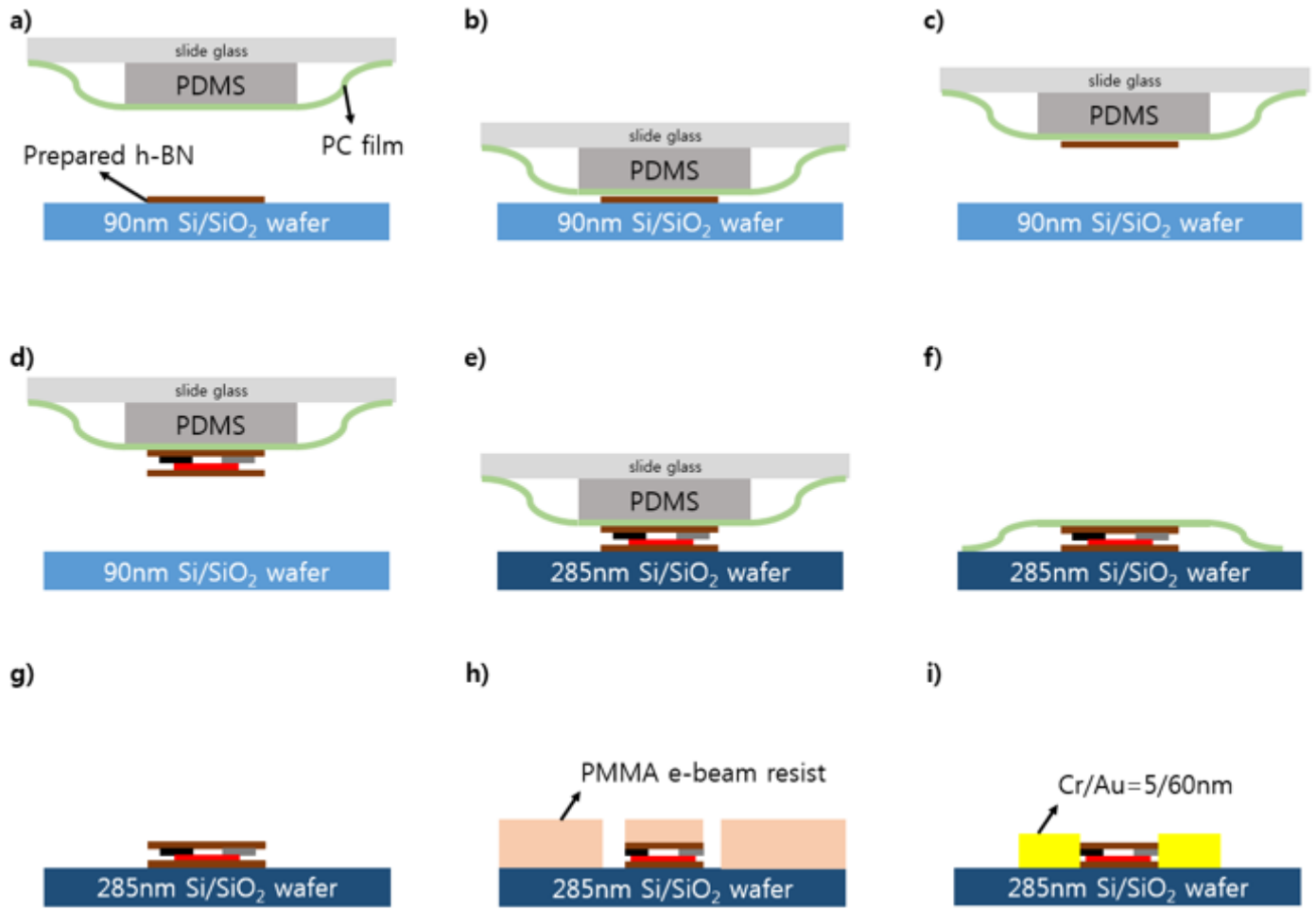

**Figure. S2 Fabrication of graphene–MoS$_2$–graphite heterojunction device: (a)** preparation of PDMS stamp covered with PC film on slide glass and mechanically exfoliated flakes (h-BN, graphene, graphite, and MoS$_2$) on 90-nm Si/SiO$_2$ wafer, **(b, c)** flake pick-up using PC film, **(d)** using the method described in **(b)** and **(c)**, pick up of graphene, graphite, MoS$_2$, and bottom h-BN in that order for heterostructure fabrication, **(e, f)** placement of stacked heterojunctions on the prepared 285-nm Si/SiO$_2$, **(g)** washing of PC film, **(h)** E-beam lithography and etching, and **(i)** evaporation of Cr/Au layer (5/60-nm-thick) and lift-off.

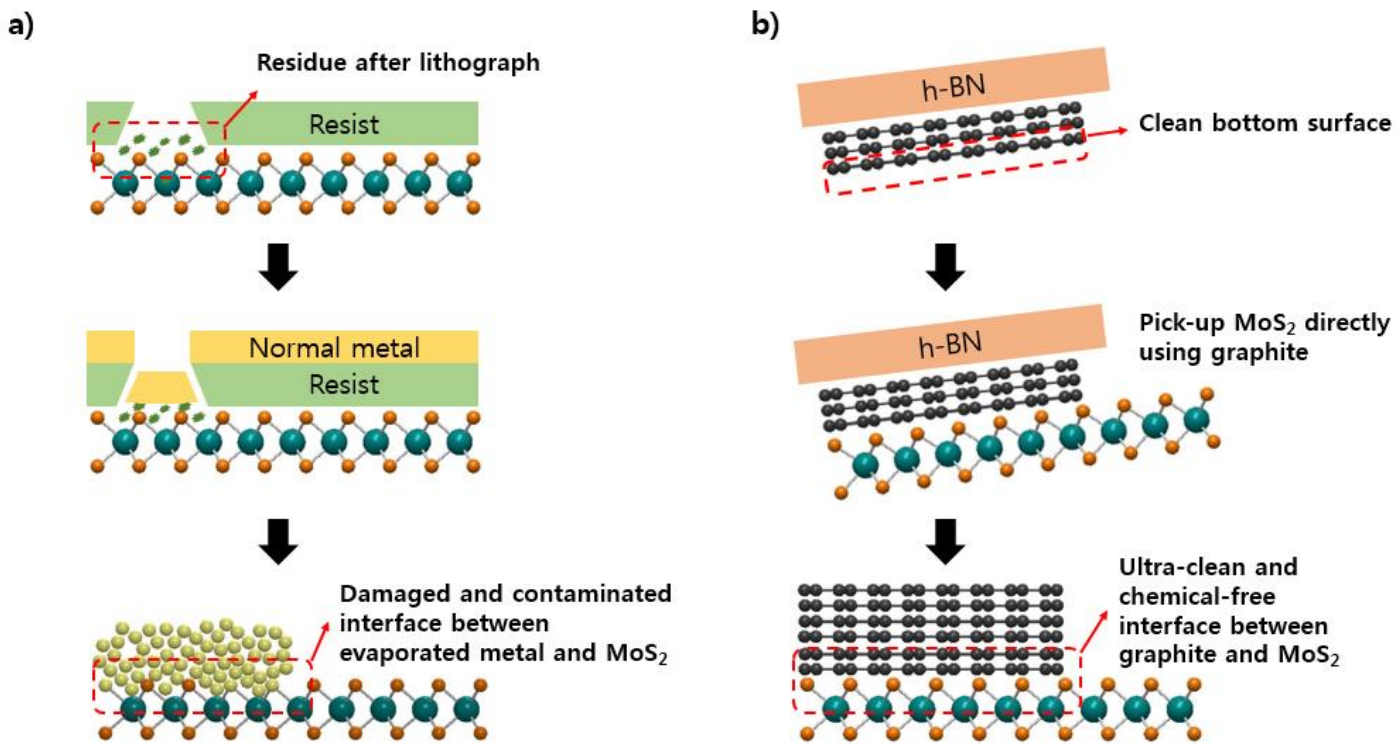

**Figure. S3 Comparison between direct metal deposition and 2D metal contacts:** (**a**) standard lithography and metal evaporation process and (**b**) 2D vdW metal contact process.

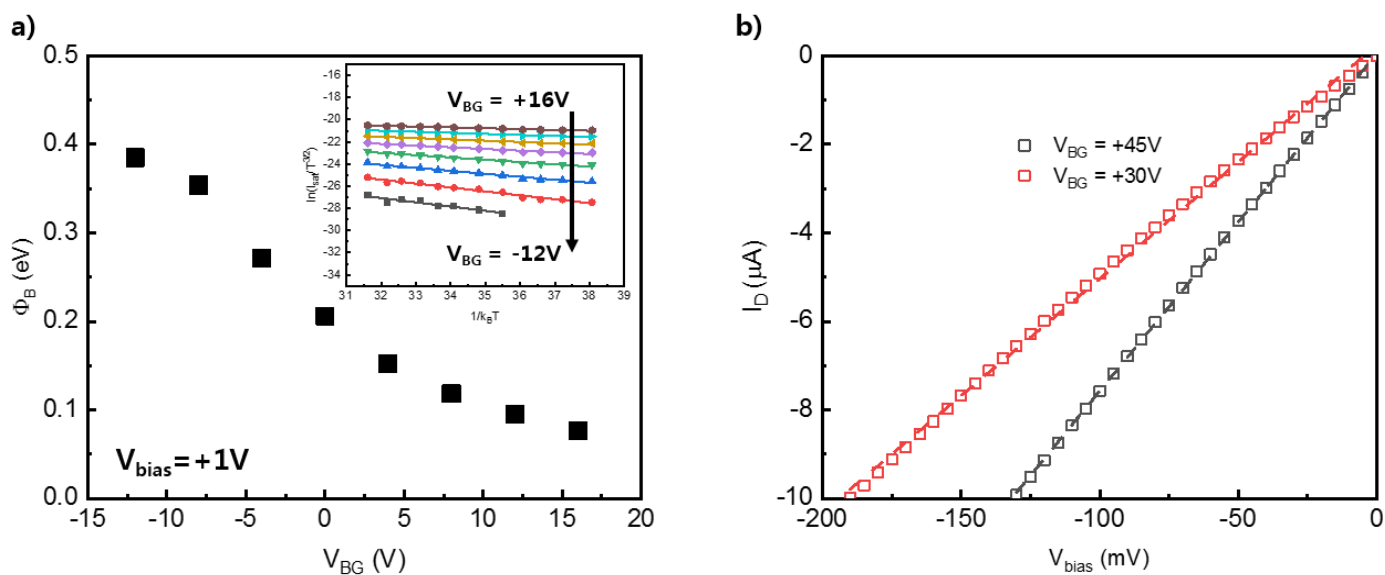

**Figure. S4** (**a**) Arrhenius plot in reverse bias regime (inset) and Schottky barrier heights obtained in this study, (**b**) current versus bias voltage characteristic in the linear non-diode regime.

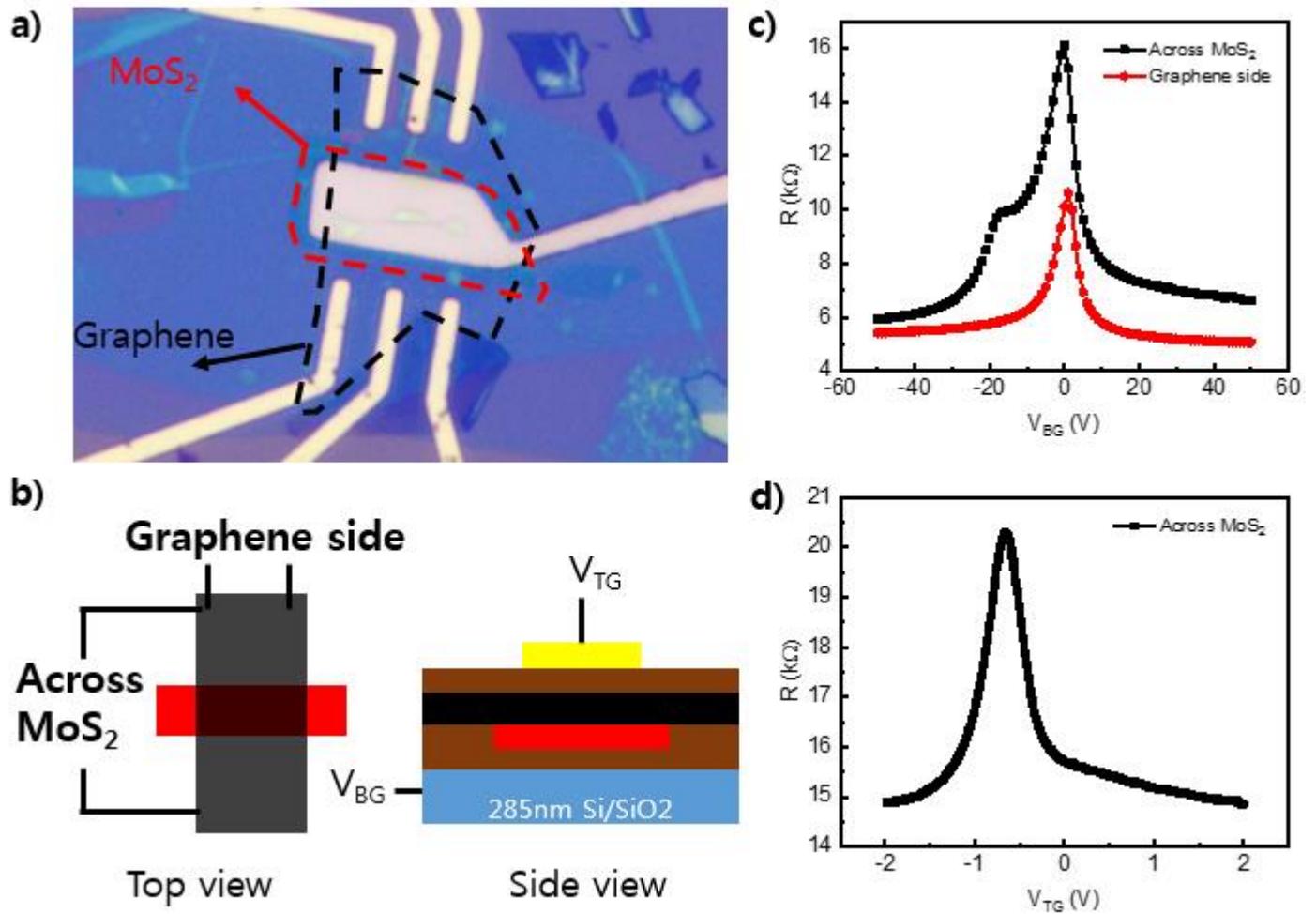

**Figure. S5 (a)** Image of the test device and **(b)** its schematic. Si acts as a global back gate, and the top gate is designed to gate the overlapping region of MoS$_2$ and graphene exclusively. In **(b)**, the black, brown, and red colors indicate graphene, hBN, and MoS$_2$ respectively. **(c)** Two-probe "across MoS$_2$" (black) and "graphene side" (red) measurements performed considering a Si back gate. **(d)** Two-probe "across MoS$_2$" measurement considering top gate.

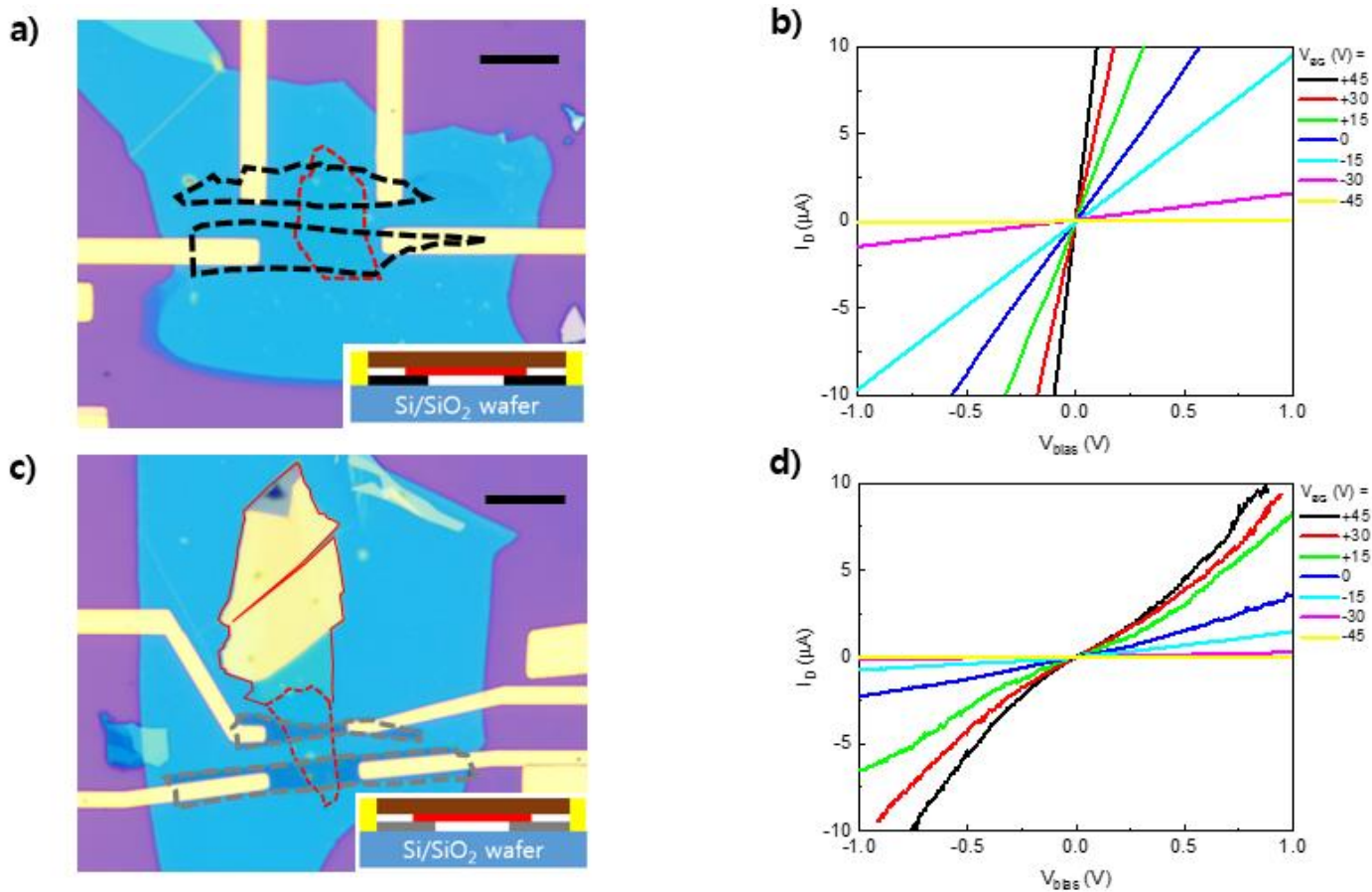

**Figure. S6 Optical images of (a) graphene–MoS$_2$–graphene and (c) graphite–MoS$_2$–graphite interfaces along with corresponding transfer curves—(b) and (d), respectively**. In **(a)** and **(c)**, the black, gray, brown, red, and yellow colors represent graphene, graphite, h-BN, MoS$_2$, and contact metal (Cr/Au), respectively. Scale bar, 10μm.

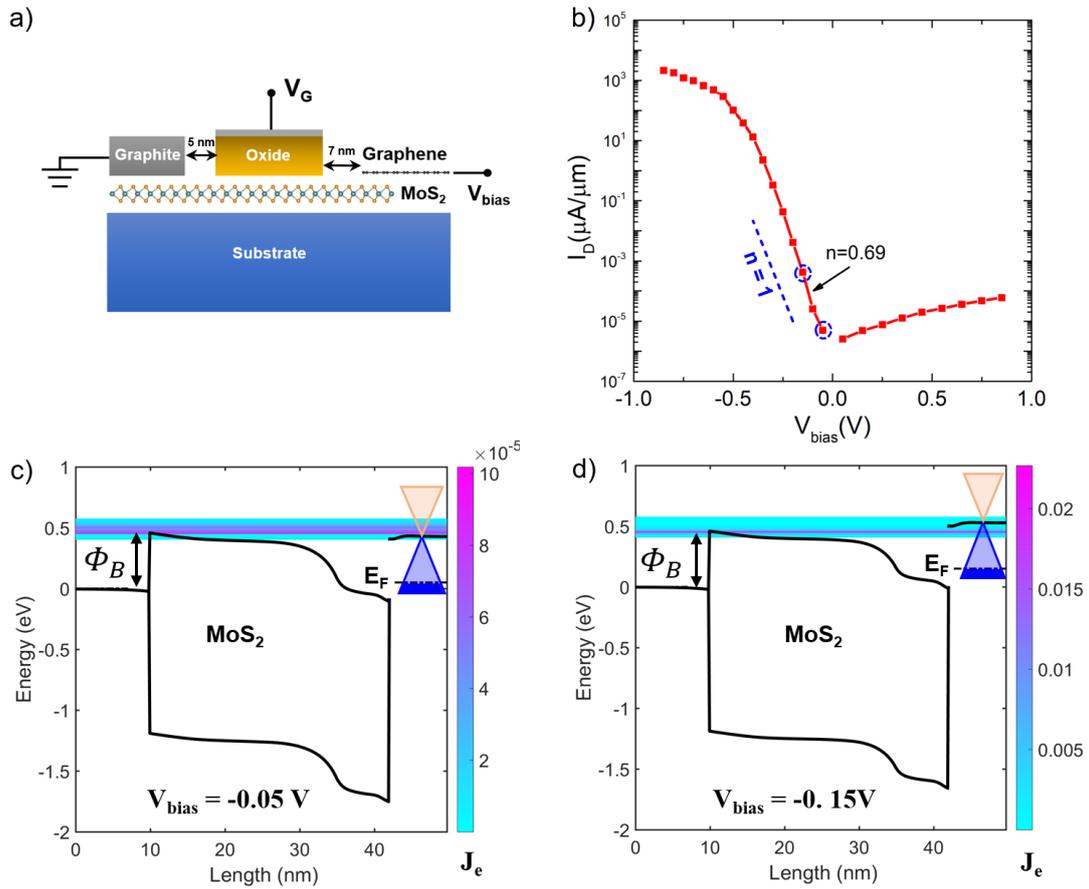

**Figure. S7** Device structure and transport properties of the simulated DS diode. (a) Schematic image of the simulated graphene–MoS$_2$ diode with 10 nm graphite, 32 nm MoS$_2$ and 8 nm graphene. Graphite-MoS$_2$ and graphene-MoS2 contacts are Schottky type and Ohmic type respectively, which is consistent with the experiment. (b) I$_D$ -V$_{bias}$ curve of the simulated DS diode. Band diagram and current density of DS diode at (c) V$_{bias}$ = -0.05 V and (d) V$_{bias}$ = -0.15 V. $\Phi_B$ is the Schottky barrier between graphite and MoS$_2$. E$_F$ is the Fermi level of graphene and the Fermi level of graphite is set at 0 eV. Unit of color bar μA/μm·eV.

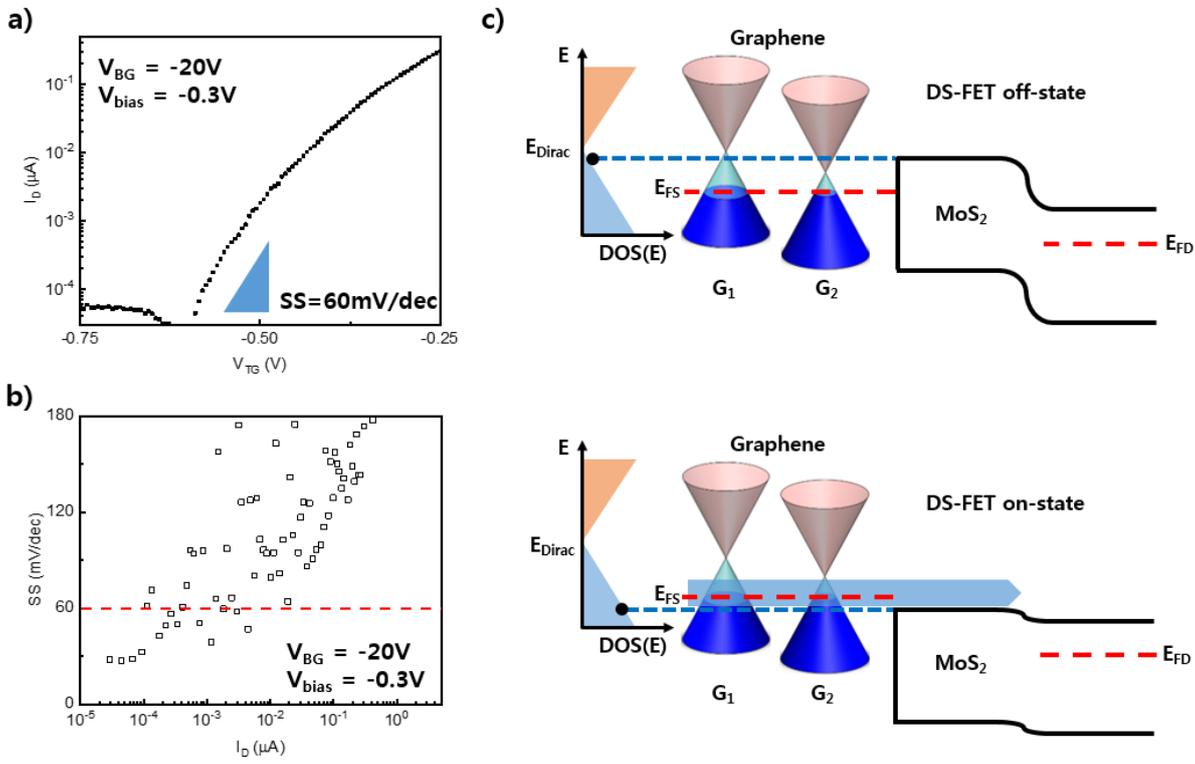

**Figure S8. MoS₂ DS-FET measurement and its band diagram. a, b** Transfer curves of the DS-FET at $V_{BG}$ = –20 V and $V_{bias}$ = –0.3 V. **c** Band diagrams of DS-FET. As the top-gate voltage, $V_{TG}$, is applied to the MoS₂ channel, the injection current density increases.

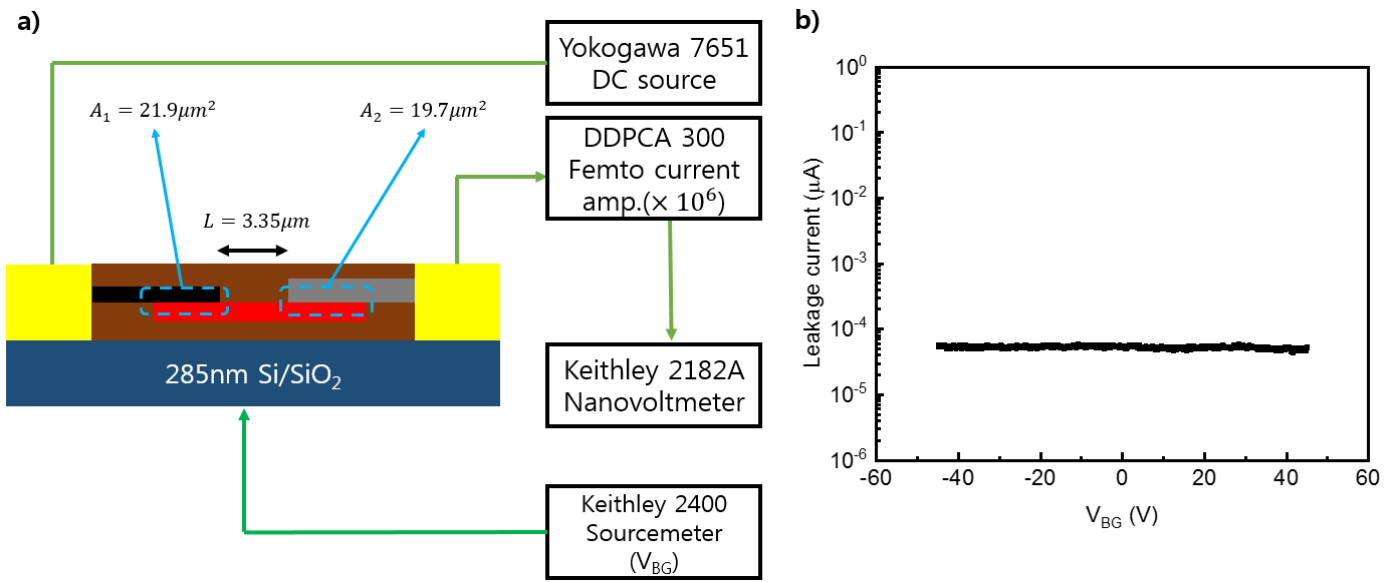

**Figure. S9 (a) Specified device dimension as well as measurement setup and (b) system leakage level observed in this study.** In **(a)**, red, black, gray, brown, and yellow colors denote MoS₂, graphene, graphite, h-BN, and contact metal (Cr/Au = 5/60-nm-thick), respectively.

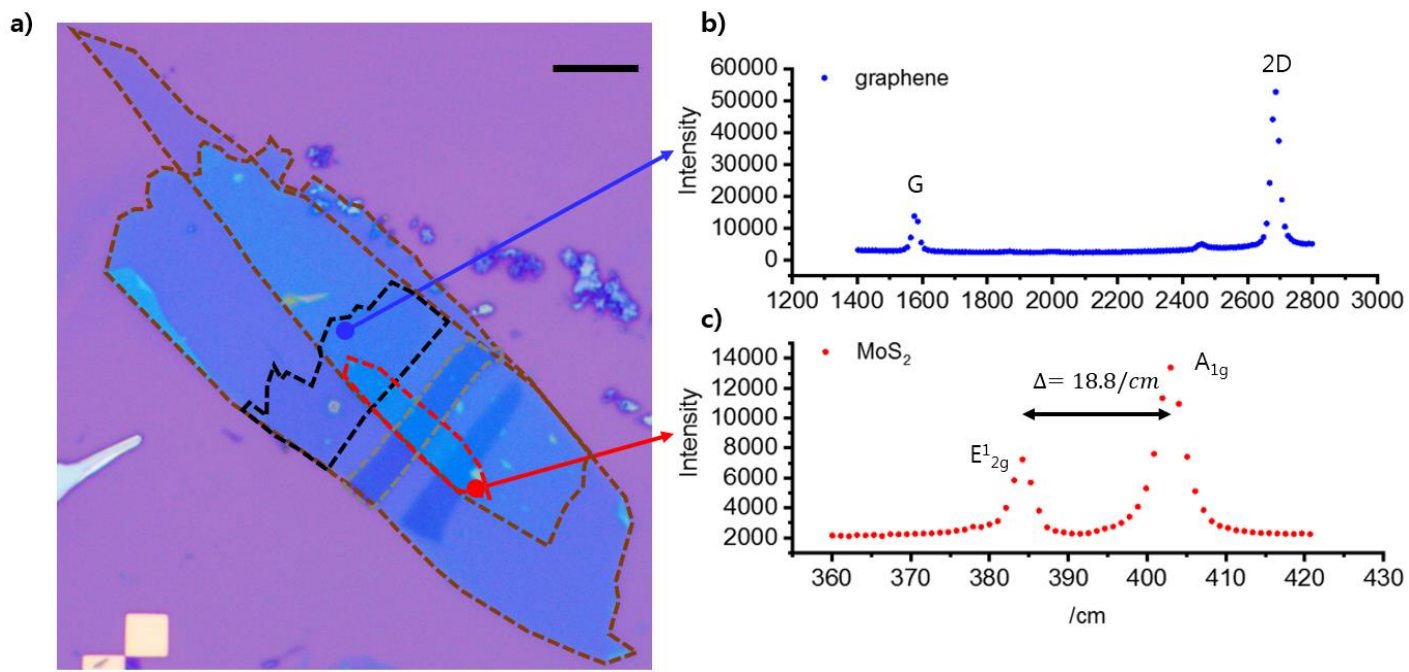

**Figure. S10 (a)** Optical image of graphene–MoS$_2$–graphite heterojunction device. The blue and red dots indicate the positions of the graphene and MoS$_2$ Raman spectra measurements, respectively. Scale bar, 10μm. **(b)** Raman spectra of graphene, and **(c)** Raman spectra of MoS$_2$.

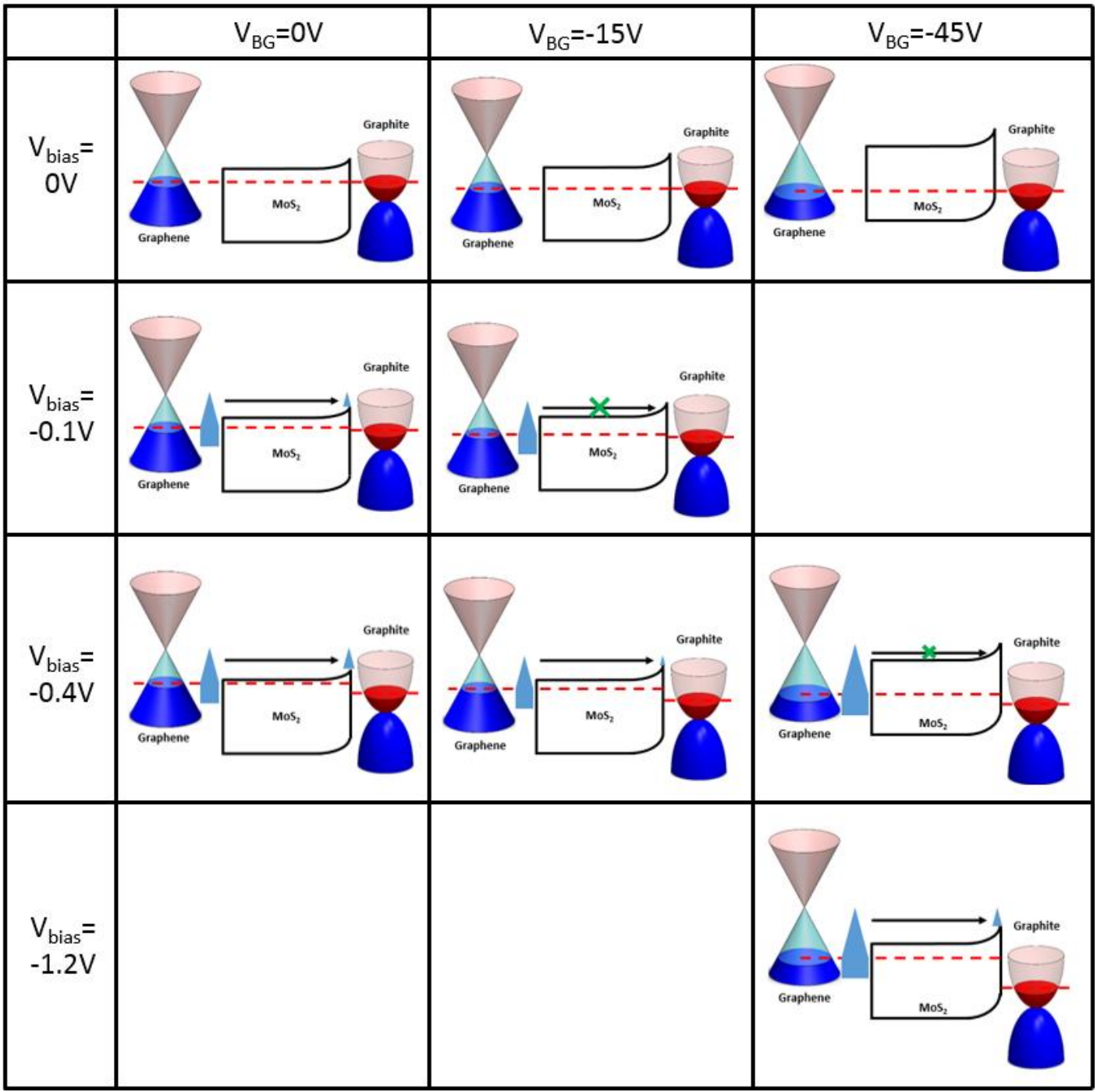

**Figure S11 Band diagrams in diode regime.**